\documentclass[aps,plb,twocolumn,showpacs,superscriptaddress,groupedaddress,nofootinbib,preprintnumbers]{revtex4-1}
\pdfoutput=1
\usepackage{amsmath}
\usepackage{amsfonts}
\usepackage{amssymb}
\usepackage{graphicx, rotating}
\usepackage{epstopdf}
\usepackage{epsfig}
\usepackage{latexsym}
\usepackage{multirow}
\usepackage{color}
\usepackage{amsmath,amssymb}
\usepackage{slashed}
\usepackage[colorlinks=true,linkcolor=red,anchorcolor=black,citecolor=blue,filecolor=cyan,menucolor=red,runcolor=filecolor,urlcolor=blue,bookmarks=true,bookmarksnumbered=true]{hyperref}
\definecolor{rossos}{cmyk}{0,1,1,0.55}
\definecolor{bluscuro}{rgb}{0.15, 0.2, .85}
\definecolor{bluchiaro}{cmyk}{1,.3,0.,0.1}

\definecolor{rossos}{cmyk}{0,1,1,0.55}
\definecolor{bluscuro}{rgb}{0.15, 0.2, .85}
\definecolor{bluchiaro}{cmyk}{1,.3,0.,0.1}

\newcommand{\bc}{\begin{center}}
\newcommand{\ec}{\end{center}}
\newcommand{\pMET}{{\bf p}\llap{/\kern1.5pt}_T}
\newcommand{\bea}{\begin{eqnarray}}
\newcommand{\eea}{\end{eqnarray}}
\newcommand{\ignore}[1]{}
\newcommand{\be}{\begin{equation}}
\newcommand{\ee}{\end{equation}}

\def\l{\label}

\def\({\left(}
\def\){\right)}
\def\<{\langle}
\def\>{\rangle}
\def\f{\frac}
\def\be{\begin{equation}}
\def\ee{\end{equation}}
\def\bry{\begin{array}}
\def\ery{\end{array}}
\def\bes{\begin{subequations}}
\def\ees{\end{subequations}}
\def\bit{\begin{itemize}}
\def\eit{\end{itemize}}
\def\ben{\begin{enumerate}}
\def\een{\end{enumerate}}

\newcommand{\MET}{E\llap{/\kern1.5pt}_T}
\definecolor{grey}{rgb}{0.6,0.6,0.6}
\definecolor{fuchsia}{rgb}{1,0,1}

\begin{document}


\title{The 750 GeV di-photon excess from the goldstino superpartner}

\author{Christoffer Petersson}
\address{Department of Fundamental Physics, Chalmers University of Technology, 412 96 G\"oteborg, Sweden}
\address{Physique Th\'eorique et Math\'ematique, Universit\'e Libre de Bruxelles, C.P. 231, 1050 Brussels, Belgium}
\address{International Solvay Institutes, Brussels, Belgium}
\author{Riccardo Torre}
\address{Institut de Th\'eorie des Ph\'enom\`enes Physiques, EPFL, CHÐ1015 Lausanne, Switzerland}
\begin{abstract}
We interpret the di-photon excess recently reported by the ATLAS and CMS collaborations as a new resonance arising from the sgoldstino scalar, which is the superpartner of the Goldstone mode of spontaneous supersymmetry breaking, the goldstino. The sgoldstino is produced at the LHC via gluon fusion and decays to photons, with interaction strengths proportional to the corresponding gaugino masses over the supersymmetry breaking scale. Fitting the excess, while evading bounds from searches in the di-jet, $Z\gamma$, $ZZ$ and $WW$ final states, selects the supersymmetry breaking scale to be a few TeV, and particular ranges for the gaugino masses. The two real scalars, corresponding to the CP-even and CP-odd parts of the complex sgoldstino, both have narrow widths, but their masses can be split of the order \mbox{10-30 GeV} by electroweak mixing corrections, which could account  
for the preference of a wider resonance width in the current low-statistics data. In the parameter space under consideration, tree-level $F$-term contributions to the Higgs mass arise, in addition to the standard $D$-term contribution proportional to the $Z$-boson mass, which can significantly enhance the tree level Higgs mass. 
\end{abstract}

\maketitle

\section{Introduction} 
\vspace{-3mm}

The ATLAS and CMS collaborations recently presented the first results based on $\sqrt{s}=13$ TeV LHC Run-II data, where both experiments showed a slight excess around 750 GeV in the di-photon invariant mass spectrum \cite{Anonymous:scEhb-ZN,ATLAS-CONF-2015-081,CMS-PAS-EXO-15-004}. The local significance of the ATLAS and CMS excesses, based on 3.2\,fb${}^{-1}$ and 2.6\,fb${}^{-1}$ of data, respectively, are 3.9$\sigma$ and 2.6$\sigma$. Several interpretations of this excess in terms of new physics have already appeared \cite{Franceschini:2015vb,Mambrini:2015wl,Angelescu:2015ua,Knapen:2015vb,Buttazzo:2015uw,Nakai:2015th,DiChiara:2015vi,Pilaftsis:2015wc,Backovic:2015tq,2015arXiv151204850H}.

In this paper we interpret the di-photon excess in terms of supersymmetry (SUSY), as arising from the sgoldstino scalar, the superpartner of the Goldstone mode of spontaneous SUSY breaking, the goldstino fermion. Since the interactions between the sgoldstino and the Standard Model (SM) particles are suppressed by the scale of SUSY breaking $\sqrt{f}$, this interpretation of the excess is only viable in SUSY models where $\sqrt{f}$ is low, of the order of a few TeV. Moreover, since these interactions are proportional to the soft masses of the superpartners, this interpretation selects particular relations and ranges for some of the superpartner masses. 

It has been previously stressed that the sgoldstino couples most strongly to SM gauge bosons, and that one of the most promising signatures is in terms of a di-photon resonance \cite{Bellazzini:2012ul,2013PhRvD..87a3008P,Dudas:2013tc}. See also refs.      
\cite{Perazzi:2000ku,
Perazzi:2000dk,
Gorbunov:2002co, Brignole:2003hb,
Antoniadis:2011ve,
Bertolini:2011wj,
Petersson:2011in,
Antoniadis:2012ui,
Dudas:2012ti,Dudas:2013tc,Astapov:2014wa, Petersson:2015te}  for different discussions concerning the sgoldstino. 
 The fact that the sgoldstino is produced at the LHC via gluon fusion implies compatibility with the $\sqrt{s}=8$ TeV LHC Run-I data, in which no significant di-photon excess was found and where a 95\% confidence level (CL) upper limit on the di-photon signal rate at around 1.5\,fb was placed \cite{ATLAScollaboration:2015uw,CMS-PAS-HIG-14-006}, since the gain in cross-section from 8 to 13\,TeV is about a factor of 4.7, in comparison to the $u\bar{u}$/$d\bar{d}$ gain of about a factor of 2.5/2.3 \cite{Franceschini:2015vb}.   
 
 The paper is organized as follows. In Section \ref{pheno} we take into account constraints arising from resonance searches in the di-jet, $Z\gamma$ and di-boson final states, and discuss the values of the SUSY breaking scale and gaugino masses relevant to explain the di-photon excess. 
In \mbox{Section \ref{Split}} we discuss the possibility of accounting for a broad width by splitting the CP-even and CP-odd part of the complex sgoldstino scalar. We discuss the implications of the sgoldstino interpretation on the Higgs sector in \mbox{Section \ref{sechiggsmass},} such as the new tree-level $F$-term contributions to the Higgs mass that it gives rise to. We conclude in \mbox{Section \ref{conclusions}.}    

\section{Explaining the di-photon excess}\label{pheno} 

In this section we will interpret the complex sgoldstino scalar \mbox{$x=(\phi+i\,a)/\sqrt{2}$}, where $\phi$ and $a$ are the CP-even and CP-odd real scalars, as being responsible for the recently reported di-photon excess. The production cross section and all the relevant partial decay widths of the sgoldstino can be found in ref.~\cite{Petersson:2015te}. Due to the experimental limit on the gluino mass and the color factor, the dominant sgoldstino partial decay width is into gluons, \mbox{$\Gamma(\phi \to gg) = (m_{3}^2 m^{3}_{\phi})/(4\pi f^{2})$}, where $m_3$ is the gluino mass, $\sqrt{f}$ is the scale of SUSY breaking and $m_{\phi}$ is the mass of $\phi$.  Thus, the sgoldstino scalars $\phi$ and $a$ are produced at the LHC via gluon fusion, with the production cross section being proportional to $\Gamma(\phi \to gg)$. The partial width into photons is instead given in terms of a linear combination of the bino and wino masses, \mbox{$\Gamma(\phi \to \gamma\gamma) = (m_{1} c_{W}^{2} +m_{2} s_{W}^{2})^{2} m^{3}_{\phi}/(32 \pi f^{2})$}, where $s_{W}$ and $c_{W}$ are the sine and cosine of the weak mixing angle. The partial decay widths of $a$ are obtained by simply replacing $\phi \to a$.

\begin{table}
\caption{\small\label{Table1} }
\begin{ruledtabular}
\begin{tabular}{c||cc}
Analysis & Constraint in units of $f$/TeV \\
\hline $jj$ \cite{ATLAScollaboration:2014el} & $ m_{3}\lesssim 0.11 $ \\
\hline $Z\gamma$ \cite{ATLAScollaboration:2014ur} & $ m_{2}{-}m_{1}  \lesssim 3.3\times 10^{-2} $ \\
\hline $ZZ$ \cite{Aad:2036291} & $ m_{1} s_{W}^{2} +m_{2} c_{W}^{2}  \lesssim 3.5\times 10^{-2} $ \\
\hline $WW$ \cite{Aad:2048180,CMScollaboration:2015ti} & $ m_2  \lesssim 4.5\times 10^{-2} $ \\
\hline $\gamma\gamma$ \cite{Anonymous:scEhb-ZN,ATLAS-CONF-2015-081,CMS-PAS-EXO-15-004} & $ 1.1\times 10^{-2} \lesssim m_{1} c_{W}^{2} +m_{2} s_{W}^{2} \lesssim 1.4\times 10^{-2} $ \\
\end{tabular}
\end{ruledtabular}\l{bounds}
\end{table}

Run-I searches for resonances in the $Z\gamma$ \cite{ATLAScollaboration:2014ur}, $ZZ$ \cite{Aad:2036291} and $WW$ \cite{Aad:2048180,CMScollaboration:2015ti} final states, place  $95\%$ CL upper limits on the signal rate at around 4\,fb, 12\,fb and 40\,fb, respectively. These constraints translate into the bounds on the sgoldstino couplings to gauge bosons, which are given in terms of ratios of different linear combinations of gaugino masses over $f$. In Table \ref{bounds} we give the constraints in terms of the sgoldstino parameter space. In the last line of the table we have translated the range preferred by the di-photon excess, obtained by requiring 6\,fb\,${<}\,\sigma\times\text{BR}_{\gamma\gamma}\,{<}\,10$\,fb at 13 TeV \cite{Franceschini:2015vb}, into a range of the relevant combination of bino and wino masses. 

\begin{figure}[t!]
\begin{center}
\includegraphics[scale=0.40]{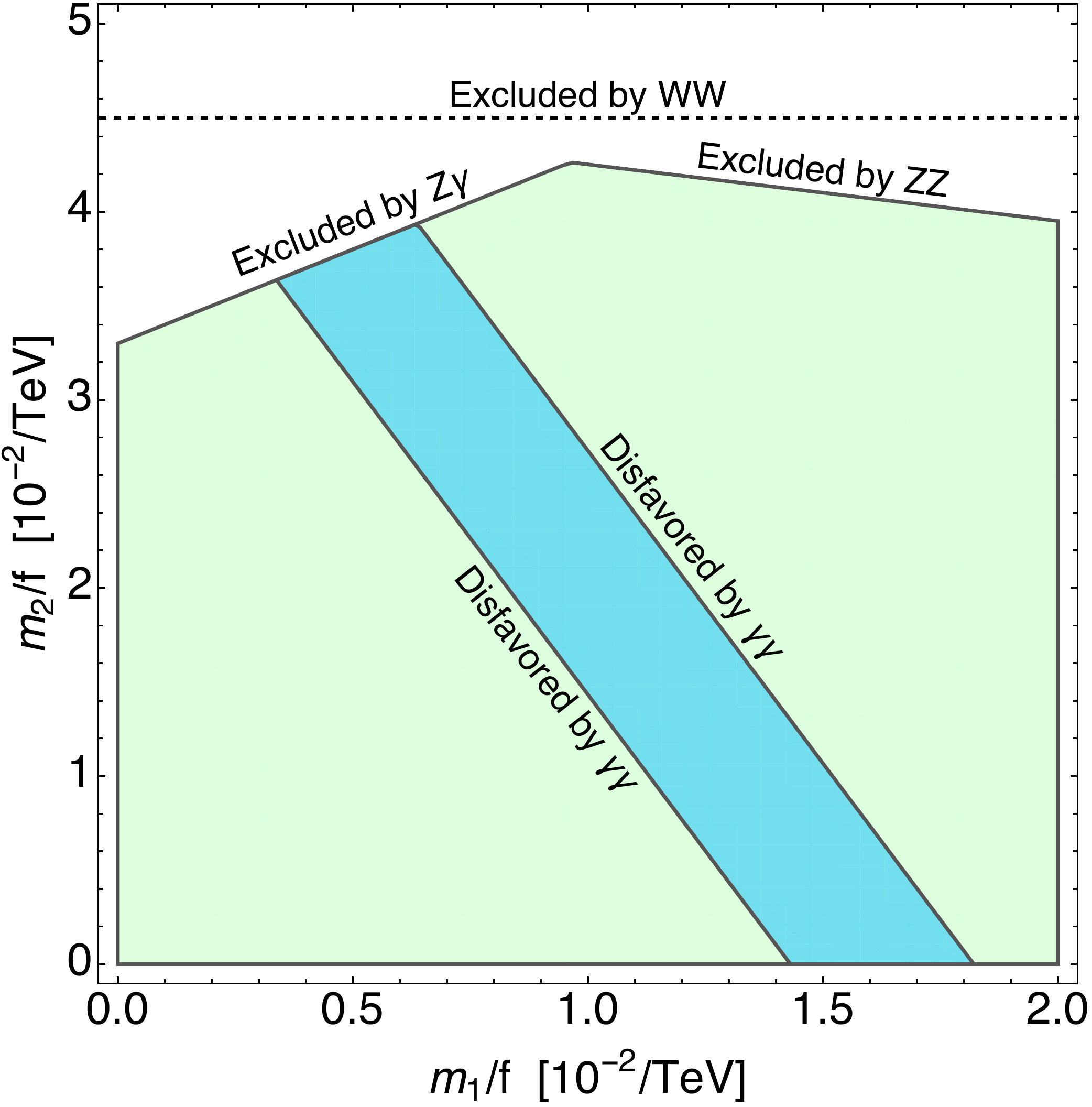}
\vspace{0.cm}\caption{The green region is allowed by the Run-I searches in the $Z\gamma$, $ZZ$ and $WW$ final states, while the blue region is  preferred by the Run-II di-photon excess. The different edges of the allowed region correspond to the exclusion limits from the $Z\gamma$ and $ZZ$ searches, respectively, while the $WW$ searches only exclude the parameter space above the dashed line.}\vspace{-7mm}
\label{fig1}
\end{center}
\end{figure}

 The interplay of the different constraints on $m_{1}$ and $m_{2}$ and the observed excess in di-photons in the plane $(m_{1}/f,m_{2}/f)$ is shown in Figure \ref{fig1}. From the figure it is clear that the strongest constraint on the region preferred by the di-photon excess come from the Run-1 $Z\gamma$ search. This implies that Run-II $Z\gamma$ searches will have great sensitivity to the sgoldstino signal hypothesis.  Note also that the constraints from the Run-I searches in the $ZZ$ and $WW$ final states are not much weaker, which suggests that, if the di-photon excess can be attributed to the sgoldstino, all the di-boson final states will show up  at about the same time.

Since the dominant decay mode of the sgoldstino is into gluons, important limits are placed by resonance searches in the di-jet final states \cite{ATLAScollaboration:2014el}.  The $95\%$ CL upper limit of 2.5\,pb on the di-jet signal rate can be translated into the bound reported in the first line of Table \ref{bounds}, which can be rewritten in the form, 
\be
\label{dijet}
\frac{\sqrt{f}}{3.9\,\text{TeV}} \gtrsim \sqrt{\frac{m_3}{1.7\,\text{TeV}}}\,,
\ee
where the constraint has been normalized to the current lower limit on the gluino mass from Run-II searches at around 1.7\,TeV \cite{Anonymous:scEhb-ZN}. For this minimum $m_{3}$ value we obtain an absolute minimum value of $\sqrt{f}$ of 3.9\,TeV. Since the maximum sgoldstino total decay width is obtained by saturating the di-jet constraint \eqref{dijet}, we conclude that the total sgoldstino width does not exceed about 0.4\,GeV. Therefore, the fact that the largest significance for the di-photon excess in the current data \cite{Anonymous:scEhb-ZN,ATLAS-CONF-2015-081,CMS-PAS-EXO-15-004} is obtained for a resonance width of around 45\,GeV can not be explained by the narrow widths of the sgoldstino scalars, if they are mass-degenerate at around 750\,GeV. In the following section we will investigate alternative explanations to account for a broader width. 

We will from hereon focus on the case where the SUSY breaking scale $\sqrt{f}$ is as low as possible, i.e.~when the di-jet constraint \eqref{dijet} is saturated. This is motivated by the fact that, as we will discuss in the Section \ref{sechiggsmass}, new $F$-term contribution to the tree level Higgs mass are maximized for low values of $\sqrt{f}$. This also maximizes the mass splitting between $\phi$ and $a$ that we propose in \mbox{Section \ref{Split}} as an explanation of the broad resonance width preferred by the data. Moreover,  low values of $\sqrt{f}$ correspond to low values of $m_3$, which is the most interesting case from the point of view of fine-tuning and gluino searches.

\begin{figure}[t!]
\begin{center}
\includegraphics[scale=0.40]{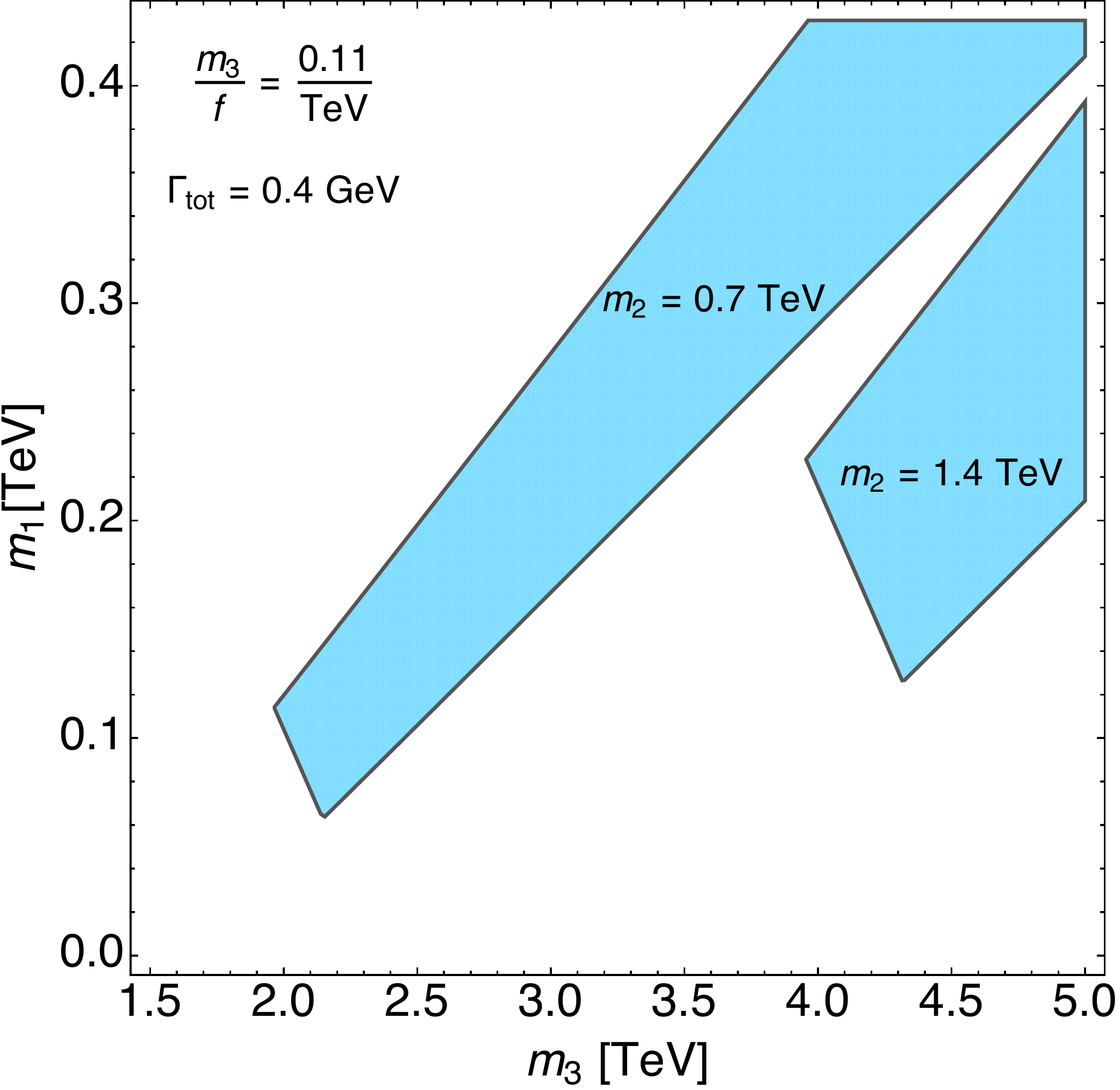}
\vspace{-0.cm}\caption{The two blue regions in the $(m_{3},m_{1})$ plane, corresponding to two different values of $m_{2}$, show the viable parameter space that can fit the di-photon excess, without being excluded by any other search channel.}\vspace{-7mm}
\label{fig2}
\end{center}
\end{figure}

In Figure \ref{fig2} we show, in the $(m_{3},m_{1})$ plane, the regions that are allowed by all Run-I constraints and where the di-photon excess can be explained by the sgoldstino. 
The two blue regions correspond to two representative values of the wino mass, $m_{2}=0.7$ and $m_{2}=1.4$ TeV. Constraints from di-jets are satisfied by construction since we require the bound in eq.~\eqref{dijet} to be saturated throughout the plane. The left edges of these regions are again due to the constraint placed by the Run-I $Z\gamma$ search \cite{ATLAScollaboration:2014ur}, while the height of the regions are determined by the di-photon signal rate preferred by the Run-II excess. 
Of course, different values of $m_{2}$ are possible and would give rise to regions that are shifted towards the left or right for smaller or larger values of $m_{2}$, respectively.

\section{Broad width from split scalars}\l{Split}

 In this section we discuss the possibility of the sgoldstino to account for the broad width around 45\,GeV for which ATLAS obtains the highest significance. We start by discussing why some possibilities, such as additional sgoldstino decays to top quarks or invisible decays, do not work  for the sgoldstino. We then discuss a more promising alternative, corresponding to splitting the masses of $\phi$ and $a$.  This latter alternative is viable because of the current low statistics in the di-photon channel, which is not yet sensitive enough to discriminate between one broad peak and two narrow peaks.
 
Concerning the explanation in terms of a large partial decay width into top quarks, this is not an option for the sgoldstino for the following reason. The sgoldstino coupling to top quarks arises from the superpotential operator corresponding to the soft $A$-term,  $(A/f)  X Q H_u U^c$, which gives rise to the following interactions,   
\be
\label{eq5}
\mathcal{L}_{ t\bar{t}} = \f{m_{t}A_t}{\sqrt{2} f} \hspace{-1pt}\left( -\phi \,t\,\bar{t} - i\,a\, t\,\gamma^5\,\bar{t}\, \right)  \, .
\ee
We see that the coupling is suppressed by the ratio $(m_{t}A_t)/ f$, implying that this decay cannot compete with the decay into gluons and therefore, cannot be responsible for a large sgoldstino width. Another option could be to enhance the sgoldstino total width by maximising the invisible width into goldstinos. However, this has the form $\Gamma(\phi \to GG) = m_{\phi}^5/(32\pi f^{2})$, which is always very suppressed for the values of $\sqrt{f}$ under consideration.

Let us now consider the possibility of splitting the masses of $\phi$ and $a$, with the purpose of generating two narrow peaks that are close by and thereby mimicking a single broad peak. The sgoldstino masses receive contributions  from the following SUSY operators
 \begin{eqnarray}
\label{xmasses}
 {\int} d^4 \theta  
 \frac{m_x^2}{4f^2} (X^\dagger X )^2 {+}\Big\{ {\int} d^2 \theta \Big( \mu {-}\f{B_\mu}{f}X \Big)  H_u H_d {+}\text{h.c.}\Big\} 
\end{eqnarray}
with $X{=}x{+}\sqrt{2}\theta \widetilde{G}{+} \theta^2 F_X$, where, in addition to the sgoldstino \mbox{$x=(\phi+i\,a)/\sqrt{2}$}, $\widetilde{G}$ is the goldstino and $F_X$ is the auxiliary field that acquires a vacuum expectation value $\langle F_X \rangle \,{ = } \,f$. The $\mu$ and $B_\mu$ parameters are the standard ones appearing in the MSSM Higgs sector. The first operator in eq.~\eqref{xmasses} provides the dominant, equal mass contribution $m_x$ to $\phi$ and $a$. However, small electroweak corrections arise from the remaining operators, which split the tree level masses of $\phi$ and $a$ according to
\be  
\label{massplit}
m_a^2 - m_{\phi}^2= \frac{2 v^2 \mu^2 B_\mu}{m_x^2 f^2} \left( 2\mu^2 \sin 2\beta - B_\mu   \right)\,,
\ee
where $v=246$\,GeV. We refer the reader to ref.~\cite{Petersson:2011in} for a treatment of all the relevant operators and the electroweak symmetry breaking conditions. 
The fact that the $\mu$ and $B_\mu$ parameters are not relevant for the di-photon excess allows for some freedom in terms of this splitting. We provide some numerical examples in the following section, where we also take into account the Higgs mass and Higgs couplings.
 
 \section{Higgs mass and couplings}\l{sechiggsmass}
 
For low values of $\sqrt{f}$, the mass of the lightest Higgs scalar $h$ receives additional tree level contributions, which arise upon integrating out the auxiliary field $F_X$ in eq.~\eqref{xmasses}, thereby generating additional quartic Higgs couplings in the $F$-term scalar potential \cite{Antoniadis:2010hs,Petersson:2011in}.  In the parameter space under consideration, the tree level Higgs mass is
\be
\label{higgsmass}
m^2_{h}\,{=}\,m_Z^2 \cos^2 2\beta \,{+} \frac{v^2}{2f^2} \Big[ \big( 2\mu^2  {-} B_\mu \sin 2\beta    \big)^2 {-}\frac{4\mu^6}{m_x^2} \sin^2 2\beta \Big]\,.
\ee
The first term is the standard MSSM $D$-term contribution. The last term arises from Higgs-sgoldstino mixing. The remaining terms arise as a consequence of treating $F_X$ dynamically, and display a destructive interference between the terms $2\mu^2$ and $B_\mu \sin 2\beta$. Note that the contribution involving $B_\mu$ is analogous to the extra tree-level contribution to the Higgs mass achieved in the NMSSM, where the role of the dimensionless coupling $\lambda$ 
is here played by the ratio $B_\mu/f$.  

An issue that is relevant to this discussion concerns the corrections to the Higgs couplings, induced by sgoldstino-Higgs mixing. As can be seen in ref.~\cite{Dudas:2012ti}, the sgoldstino-Higgs mixing corrections to the Higgs coupling to gluons, photons and $Z\gamma$ are all proportional to $v^2 \mu^3 \sin 2\beta/(m_x^2 f^2)$. Moreover, the different corrections are proportional to the corresponding linear combination of gaugino masses, which can be found in Table~\ref{Table1}. The fact that the Higgs corrections depend cubically on $\mu$ severely constrains the possibility of using large values of $\mu$ to get a substantial mass splitting \eqref{massplit} and an enhanced tree level Higgs mass \eqref{higgsmass}, even for large values of $\tan\beta$. We find that, by requiring not more than 10\% modification to the Higgs couplings, we can neither achieve the mass splitting needed to explain the broad width nor get a  significant Higgs mass enhancement at tree level. 

Instead, the only viable possibility is to consider small values of $\tan\beta$ and large values of $B_\mu$, thereby maximizing the contribution from the $B_\mu \sin 2\beta$ term in eq.~\eqref{higgsmass}. In order to minimize the cancellation in eq.~\eqref{higgsmass}, small values of $\mu$ are now required, implying that the Higgs coupling corrections, as well as the last term in eq.~\eqref{higgsmass}, both of which arise from sgoldstino-Higgs mixing, are small. 
As a numerical example, for $B_\mu/f=0.8$, $\mu= 400$ GeV and $\tan\beta=2$, one obtains a tree level Higgs mass around $m_{h}=120$\,GeV and a mass splitting between $\phi$ and $a$ of about $15$ GeV, while keeping the modifications to the Higgs couplings below 10\%. 


\section{Conclusions}\l{conclusions}
Sharing the excitement of the theory community for the recent announcement of an excess of events in the di-photon spectrum at an invariant mass of about $750$ GeV \cite{Anonymous:scEhb-ZN,ATLAS-CONF-2015-081,CMS-PAS-EXO-15-004}, we propose an interpretation in terms of the complex scalar superpartner of the goldstino, the sgoldstino. We have studied  the parameter space where this interpretation is compatible with the excess, while evading all other constraints. The production cross section and branching ratios of the sgoldstino  depend  only on the gaugino masses and the SUSY breaking scale, and the strong limits from Run-I resonance searches set strong constraints on them. Nevertheless we find an allowed region of the parameter space pointing towards hierarchical gaugino masses $m_{1}<m_{2}<m_{3}$ and a low SUSY breaking scale in the few TeV range. 

We also studied the possibility to, within the sgoldstino interpretation, mimic the large width of around 45\,GeV, as suggested by data.  Since the dominant sgoldstino decay width is into two gluons, the constraints from di-jet searches in Run-I set the maximum allowed width to less than a GeV. However, a natural splitting between the masses of the CP-even and CP-odd real sgoldstino scalars arises from electroweak mixing corrections, and can be in the range $10-30$ GeV. This would allow for an explanation of a broader peak, while at the same time providing a significant  additional $F$-term tree level contribution to the tree level Higgs mass. Due to the small width of the two scalars and to the good experimental invariant mass resolution we expect that the two peaks could be resolved by the experiments with a slight more statistics. 

Let us stress that if the di-photon excess is due to the sgoldstino scalar, this would provide crucial information about the full supersymmetric model that lies beyond the SM, as if would select a range for the SUSY breaking scale that is lower than the typical range selected by the standard SUSY frameworks such as gauge mediation, gravity mediation and anomaly mediation. Another interesting aspect of the sgoldstino interpretation is  the fact that it predicts relations between seemingly disconnected experimental analyses, such as direct searches for gluinos, winos, higgsinos as well as Higgs measurements and searches for new resonances. And given the relations it predicts between different  gauge boson channels, we expect that hints should appear in the di-jet, $Z\gamma$ and $ZZ$ channels already with the next few inverse femtobarns of data. 
We are looking forward to knowing whether this signal is actually due to new physics or yet another statistical fluctuation.  

\vspace{0.5cm}
\noindent {\bf Acknowledgments:}
 The work of C.\,P.~is supported by the Swedish Research Council (VR) under the contract 637-2013-475, by IISN-Belgium (conventions 4.4511.06, 4.4505.86 and 4.4514.08) and by the ``Communaut\'e Fran\c{c}aise de Belgique" through the ARC program and by a ``Mandat d'Impulsion Scientifique" of the F.R.S.-FNRS. 
The work of RT is supported by Swiss National Science Foundation under grants CRSII2-160814 and
200020-150060.


\bibliographystyle{mine}
\bibliography{bibliography}

\end{document}